\documentclass[12pt]{article}
\input tcilatex
\QQQ{Language}{
American English
}

\begin{document}

\title{An elliptic inequality for nonlinear Hodge fields}
\author{Thomas H. Otway \\
%EndAName
\\
\textit{Departments of Mathematics and Physics,}\\
\textit{Yeshiva University, New York, NY 10033}}
\date{}
\maketitle

\begin{abstract}
A version of the nonlinear Hodge equations is introduced in which the
irrotationality condition is weakened. An elliptic estimate for solutions is
derived. 1991 MSC: 58E15 (Classical field theory)
\end{abstract}

\section{\textrm{Introduction}}

In this note we study systems of the form 
\begin{equation}
\delta \left( \rho (Q)\omega \right) =0,
\end{equation}
\begin{equation}
d\omega =u\wedge \omega ,
\end{equation}
where $\omega \in \Lambda ^p\left( T^{*}M\right) $ for $p\geq 1;$\textit{\ }$%
u\in \Lambda ^1\left( T^{*}M\right) ;$ \textit{M }is an \textit{n-}%
dimensional Riemannian manifold; $\delta :\Lambda ^p\rightarrow \Lambda
^{p-1}$ is the adjoint of the exterior derivative $d;\;Q=\left\langle \omega
,\omega \right\rangle \equiv *(\omega \wedge *\omega );$ $*:\Lambda
^p\rightarrow \Lambda ^{n-p}$ is the Hodge involution; $\rho :\mathbf{R}%
^{+}\rightarrow \mathbf{R}^{+}$ is a $C^1$ function satisfying the condition 
\textbf{[U]} 
\begin{equation}
K^{-1}(Q+k)^q\leq \rho (Q)+2Q\rho ^{\prime }(Q)\;\leq K(Q+k)^q
\end{equation}
for some positive constant $K\;$and nonnegative constants $k,q.$

If $u\equiv 0$ (or if $p=1$ and $u=\omega $), then condition (2) degenerates
to the condition 
\begin{equation}
d\omega =0.
\end{equation}
Equation (1) under condition (4) was introduced and extensively studied by
L. M. and R. J. Sibner (\textbf{[SS1]}-\textbf{[SS4]}).

Equation (1) arises in a variety of physical and geometric contexts. For
example, if we choose 
\begin{equation}
\rho (Q)=\left( 1-\frac{\gamma -1}2Q\right) ^{1/(\gamma -1)},
\end{equation}
then for $p=1,$ eq. (1) describes the velocity field of a stationary,
polytropic, compressible flow, where $\gamma >1$ is the adiabatic constant
of the fluid \textbf{[SS1]}. Condition (3) with $q=0$ is a condition for
subsonic flow. For a suitable choice of $\rho ,$ eq. (1) can also be used to
model nonparametric minimal surfaces of codimension 1 \textbf{[SS2]}. If $%
p=2,$ then eq. (1) can be used to model the external magnetic field of a
body with magnetic permeability $\rho ^{-1}$ (see, \textit{e.g.,} \textbf{%
[O2])}.

Condition (4) guarantees, via the converse of Poincar\'{e}'s Lemma, the
local existence of a \textit{potential:} a $p-1$-form $\varphi $ such that $%
\omega =d\varphi .$ If $\omega \in \Lambda ^1(T^{*}M)$ is the velocity field
of an \textit{n}-dimensional fluid, then the multivalued nature of the
0-form $\varphi $ corresponds to circulation about handles in a nontrivial
topology. Condition (4) guarantees that the flow is \textit{irrotational:}
no circulation exists about any curve homologous to zero.

If $\omega \in \Lambda ^1(T^{*}M),$ then condition (2) only guarantees, via
the Frobenius Theorem, that $\omega =\ell d\varphi $ locally; a potential
exists only along the hypersurfaces $\ell =$ \textit{constant,} and
circulation about topologically trivial points is excluded only along these
hypersurfaces. (For the extension of this result to exterior products of
1-forms, see, \textit{e.g.,} Sec. 4-3 of \textbf{[E]}.)

We have as an immediate result of (2) the condition 
\begin{equation}
d\omega \wedge \omega =0.
\end{equation}
If $\omega $ denotes tangential velocity of a rigid rotor ($\rho =\rho (x)$
only), eq. (6) corresponds in three euclidean dimensions to the fact that
the direction of $\nabla \times \omega $ is perpendicular to the plane of
rotation. Condition (6) also arises in thermodynamics \textbf{[C]}, \textbf{%
[E]}.

The replacement of condition (4) by condition (2) is mathematically
significant for the following reasons. Regarding the existence of solutions
and their uniqueness, eqs. (4) can be used to prescribe a cohomology class
for solutions, whereas eq. (2) will only prescribe a closed ideal. An $%
L^\infty $ bound on weak solutions to (1) follows, via a Morrey estimate,
from a subelliptic inequality for the scalar $Q.$ The existing estimates (%
\textbf{[U]},\textbf{\ [SS3]}, \textbf{[SS4]}) all rely on (4). Additional
remarks on applications are given at the end of Sec. 2.

\section{Elliptic estimate}

\begin{theorem}
Let $\omega ,u$ smoothly satisfy eqs. (1), (2) on a bounded, open domain $%
\Omega \subset \mathbf{R}^n.$ Assume condition (3). Then the scalar $%
Q=*(\omega \wedge *\omega )$ satisfies the elliptic inequality 
\begin{equation}
L_\omega (Q)+C(Q+k)^q(|\nabla u|+|u|^2)Q\geq 0,
\end{equation}
where $L_\omega $ is a divergence-form operator which is uniformly elliptic
for $k>0.$ (Here and throughout, we denote by $C$ generic positive constants
the value of which may change from line to line.)
\end{theorem}

\textit{Proof.} We have (\textbf{[U]}, (1.5)-(1.7)) 
\[
\left\langle \omega ,\Delta \left( \rho (Q)\omega \right) \right\rangle
=\partial _i\left\langle \omega ,\partial _i\left( \rho (Q)\omega \right)
\right\rangle -\left\langle \partial _i\omega ,\partial _i\left( \rho
(Q)\omega \right) \right\rangle 
\]
\begin{equation}
=\Delta H(Q)-\left[ \rho (Q)\left\langle \partial _i\omega ,\partial
_i\omega \right\rangle +\rho ^{\prime }(Q)\left\langle \partial _i\omega
,\omega \right\rangle \partial _iQ\right] ,
\end{equation}
where 
\[
\Delta H(Q)=\partial _i\left[ \left( \frac 12\rho (Q)+Q\rho ^{\prime
}(Q)\right) \partial _iQ\right] ,
\]
$\partial _i=\partial /\partial x^i,x=x^1,...,x^n\in \Omega ,$ and, here and
throughout, we use the Einstein convention for repeated indices. Observe
that $H$ is defined so that 
\[
H^{\prime }(Q)=\frac 12\rho (Q)+Q\rho ^{\prime }(Q).
\]
Moreover 
\begin{equation}
\rho ^{\prime }(Q)\left\langle \partial _i\omega ,\omega \right\rangle
\partial _iQ=\sum_i2\rho ^{\prime }(Q)\left\langle \partial _i\omega ,\omega
\right\rangle ^2.
\end{equation}
If $\rho ^{\prime }(Q)\geq 0,$ then (9) implies that 
\begin{equation}
\rho (Q)\left\langle \partial _i\omega ,\partial _i\omega \right\rangle
+\rho ^{\prime }(Q)\left\langle \partial _i\omega ,\omega \right\rangle
\partial _iQ\geq \rho (Q)\left| \nabla \omega \right| ^2\geq
K^{-1}(Q+k)^q\left| \nabla \omega \right| ^2.
\end{equation}
In (10) we have used the inequality 
\begin{equation}
\rho (Q)\geq K^{-1}(Q+k)^q,
\end{equation}
which follows from (3) (with a possibly larger constant $K)$. If $\rho
^{\prime }(Q)<0,$ then (9) and the Schwarz inequality imply that 
\[
\rho (Q)\left\langle \partial _i\omega ,\partial _i\omega \right\rangle
+\rho ^{\prime }(Q)\left\langle \partial _i\omega ,\omega \right\rangle
\partial _iQ\geq \rho (Q)\left| \nabla \omega \right| ^2+2\rho ^{\prime
}(Q)\left| \nabla \omega \right| ^2Q=
\]
\begin{equation}
\left[ \rho (Q)+2Q\rho ^{\prime }(Q)\right] \left| \nabla \omega \right|
^2\geq K^{-1}(Q+k)^q\left| \nabla \omega \right| ^2.
\end{equation}
Thus (8) implies, via either (10) or (12) as appropriate, the inequality 
\begin{equation}
\left\langle \omega ,\Delta \left( \rho (Q)\omega \right) \right\rangle \leq
\Delta H(Q)-K^{-1}(Q+k)^q\left| \nabla \omega \right| ^2.
\end{equation}
Applying eq. (1) to the left-hand side of (13) yields, for $\Delta \equiv
-\left( d\delta +\delta d\right) ,$%
\[
\left\langle \omega ,\Delta \left( \rho (Q)\omega \right) \right\rangle
=-*\left[ \omega \wedge *\delta d\left( \rho (Q)\omega \right) \right]
=(-1)^{n(p+1)+n}*\left[ \omega \wedge *(*d*)d(\rho \omega )\right] 
\]
\[
=(-1)^{n(p+1)+n+n(n-p)+n-p}*\left[ \omega \wedge d*d(\rho \omega )\right] =
\]
\[
(-1)^{n(p+1)+n+n(n-p)+n-p+p}*\left\{ d\left[ \omega \wedge *d(\rho \omega
)\right] -\left[ d\omega \wedge *d(\rho \omega )\right] \right\} 
\]
\begin{equation}
=(-1)^{n(n+3)}\left\{ *d\left[ \omega \wedge *d(\rho \omega )\right]
-*\left[ u\wedge \omega \wedge *d(\rho \omega )\right] \right\} \equiv \tau
_1-\tau _2.
\end{equation}
We express the first term in this difference, up to sign, as a divergence in
the 1-form $dQ,$writing 
\[
\tau _1=*d\left[ \omega \wedge *d(\rho \omega )\right] =
\]
\begin{equation}
\ast d\left[ \omega \wedge *\left( \rho ^{\prime }(Q)dQ\wedge \omega \right)
\right] +*d\left[ \omega \wedge *\rho d\omega \right] \equiv \tau _{11}+\tau
_{12}.
\end{equation}
Notice that 
\[
\ast d\alpha =(-1)^n\delta *\alpha =(-1)^ndiv\,(*\alpha )
\]
for $\alpha \in \Lambda ^{n-1}.$ Equation (2) implies that 
\[
\tau _{12}\geq -|\tau _{12}|=-\left| *d\left[ \omega \wedge *(\rho u\wedge
\omega )\right] \right| \geq 
\]
\begin{equation}
-C\left( |\nabla \omega ||u|\rho |\omega |+|\nabla u|\rho Q+|u||\omega
||\nabla (\rho \omega )|\right) \equiv C\left( -\tau _{121}-\tau _{122}-\tau
_{123}\right) .
\end{equation}
We have, analogously to (11), the inequality $\rho (Q)\leq K(Q+k)^q.$ Using
this estimate and Young's inequality, we write 
\begin{equation}
-\tau _{121}=-\sqrt{\rho }|\nabla \omega ||u|\sqrt{\rho }|\omega |\geq
-\varepsilon |\nabla \omega |^2(Q+k)^q-C(\varepsilon ,K)|u|^2(Q+k)^qQ.
\end{equation}
Kato's inequality and (3) yield, using $\left| \rho ^{\prime }(Q)\cdot
Q\right| \leq K(Q+k)^q,$ 
\[
-\tau _{123}=-|u||\omega ||\nabla (\rho \omega )|=-|u||\omega ||\rho
^{\prime }(Q)\nabla Q\cdot \omega +\rho \nabla \omega |\geq 
\]
\[
-|u||\omega |\left( \left| 2\rho ^{\prime }(Q)|\omega |\nabla |\omega |\cdot
\omega \right| +\left| \rho (Q)\nabla \omega \right| \right) \geq 
\]
\[
-2|u||\omega |\left| \rho ^{\prime }(Q)\cdot Q\right| \left| \nabla |\omega
|\right| -|u||\omega |K(Q+k)^q\left| \nabla \omega \right| 
\]
\begin{equation}
\geq -3|u||\omega |K(Q+k)^q|\nabla \omega |\geq -K(Q+k)^q\left( \varepsilon
|\nabla \omega |^2+C(\varepsilon )|u|^2Q\right) .
\end{equation}
Substituting (17) and (18) into (16) yields, for a new $\varepsilon $, 
\begin{equation}
\tau _{12}\geq -|\tau _{12}|\geq -K\varepsilon (Q+k)^q|\nabla \omega
|^2-\left( C(\varepsilon ,K)|u|^2+K|\nabla u|\right) (Q+k)^qQ.
\end{equation}
Similarly, 
\[
\tau _2=*\left[ u\wedge \omega \wedge *d(\rho \omega )\right] \geq
-C|u||\omega ||\nabla (\rho \omega )|,
\]
which can be estimated by (18). Substituting (19) into (15), (15) into (14),
and (14) into (13), and estimating $\tau _2$ of (14) by (18) yields, again
for a new $\varepsilon $, 
\[
\ast d\left[ \omega \wedge *\left( \rho ^{\prime }(Q)dQ\wedge \omega \right)
\right] -K\varepsilon (Q+k)^q|\nabla \omega |^2
\]
\[
-C(\varepsilon ,K)(Q+k)^q\left( |\nabla u|+|u|^2\right) Q\leq \Delta
H(Q)-K^{-1}(Q+k)^q|\nabla \omega |^2.
\]
We obtain, choosing $0<\varepsilon \leq K^{-2}$ , 
\[
0\leq (K^{-1}-\varepsilon K)(Q+k)^q|\nabla \omega |^2\leq \Delta H(Q)\pm
div\left( *\left[ \omega \wedge *\left( \rho ^{\prime }(Q)dQ\wedge \omega
\right) \right] \right) 
\]
\[
+C(Q+k)^q\left( |\nabla u|+|u|^2\right) Q\equiv L_\omega (Q)+C(Q+k)^q\left(
|\nabla u|+|u|^2\right) Q.
\]

The ellipticity of the operator $L_\omega $ under condition (3) is obvious
from the definition of $\Delta H,$ for either choice of sign in the other
second-order term. Ellipticity can also be recovered from the arguments of [%
\textbf{U}], Section 1, as $L_\omega $ includes no terms arising from the
right-hand side of (2). This completes the proof of the theorem. 
\[
\]

We can derive some minor improvements to the literature of nonlinear Hodge
theory on the basis of Theorem 1, beyond the weakening of condition (4).
Lemma 2.3 of \textbf{[S]} and Theorem 3.1 of \textbf{[O2]} assume, in
addition to (4), that $q=0$ in (3). Theorem 1 implies obvious extensions of
those results. (The precise form of the extension of [\textbf{O2}], Theorem
3.1, depends on the degree of smoothness that one is willing to ascribe to $%
u.)$ Moreover, our arguments imply that the coefficients $b^{\,j}$ in
expression (2.2) of \textbf{[S]} are zero, at least locally, and that the
elliptic operator in Proposition 1.2 of \textbf{[SS4]} is a divergence-form
operator on $Q$. The arguments of this section improve and extend the
corresponding estimates of [\textbf{O1}], Section 3, and \textbf{[O2]},
Section 4. Details are given in [\textbf{O3}].

\[
\]

\[
\mathbf{References} 
\]

\textbf{[C]} C. Carath\'{e}odory, Gesammelte Mathematische Schriften, Bd.
II, S. 131-177, C. H. Beck'sche Verlagsbuchhandlung, M\"{u}nich, 1955.

\textbf{[E]} D. G. B. Edelen, Applied Exterior Calculus, Wiley, New York,
1985.

\textbf{[O1]} T. H. Otway, Yang-Mills fields with nonquadratic energy, 
\textit{J. Geometry \& Physics} \textbf{19 }(1996),379-398.

\textbf{[O2]} T. H. Otway, Properties of nonlinear Hodge fields, \textit{J.
Geometry \& Physics,} in press.

[\textbf{O3}] T. H. Otway, Nonlinear Hodge equations in vector bundles,
preprint.

\textbf{[S]} L. M. Sibner, An existence theorem for a nonregular variational
problem, \textit{Manuscripta} \textit{Math.} \textbf{43 }(1983), 45-72.

\textbf{[SS1]} L. M. Sibner and R. J. Sibner, A nonlinear Hodge-de Rham
theorem, \textit{Acta Math.} \textbf{125 }(1970), 57-73.

\textbf{[SS2]} L. M. Sibner and R. J. Sibner, Nonlinear Hodge theory:
Applications, \textit{Advances in Math.} \textbf{31} (1979), 1-15.

\textbf{[SS3]} L. M. Sibner and R. J. Sibner, A maximum principle for
compressible flow on a surface, \textit{Proc. Amer. Math. Soc.} \textbf{71}%
(1) (1978), 103-108.

\textbf{[SS4]} L. M. Sibner and R. J. Sibner, A subelliptic estimate for a
class of invariantly defined elliptic systems, \textit{Pacific J. Math.} 
\textbf{94}(2) (1981), 417-421.

\textbf{[U]} K. Uhlenbeck, Regularity for a class of nonlinear elliptic
systems, \textit{Acta Math.} \textbf{138 }(1977), 219-240.

\end{document}